\documentclass[twocolumn,aps,prl,amssymb,amstext,showpacs]{revtex4}

\usepackage{amsfonts,amssymb,amsmath,graphicx}
\pdfoutput=1

\newcommand{\ud}{\mathrm{d}}
\newcommand{\sgn}{\mathrm{sgn}}
\newcommand{\erf}{\mathrm{erf}}

\begin{document}

\title{Diffraction in Time: An Exactly Solvable Model}

\author{Arseni Goussev}

\affiliation{Department of Mathematics and Information Sciences,
  Northumbria University, Newcastle Upon Tyne, NE1 8ST, United
  Kingdom}

\affiliation{Max Planck Institute for the Physics of Complex Systems,
  N{\"o}thnitzer Stra{\ss}e 38, D-01187 Dresden, Germany}

\date{\today}

\begin{abstract}
  In the recent years, mater-wave interferometry has attracted growing
  attention due to its unique suitability for high-precision
  measurements and study of fundamental aspects of quantum
  theory. Diffraction and interference of matter waves can be observed
  not only at a spatial aperture (such as a screen edge, slit, or
  grating), but also at a time-domain aperture (such as an absorbing
  barrier, or ``shutter'', that is being periodically switched on and
  off). The wave phenomenon of the latter type is commonly referred to
  as ``diffraction in time''. Here, we introduce a versatile, exactly
  solvable model of diffraction in time. It describes time-evolution
  of an arbitrary initial quantum state in the presence of a
  time-dependent absorbing barrier, governed by an arbitrary aperture
  function. Our results enable a quantitative description of
  diffraction and interference patterns in a large variety of setups,
  and may be used to devise new diffraction and interference
  experiments with atoms and molecules.
\end{abstract}


\maketitle

%%%%%%%%%%%%%%%%%%%%%%%%%%%%%%%%%%%%%%%%%%%%%%%%%%%%%%%%%%%%%%%%%%%%%%%%%%%%%%%%
The wave nature of matter is both captivating and perplexing, and its
exploration has been at the center of experimental and theoretical
research since early days of quantum mechanics. To date, diffraction
and interference experiments have been successfully performed with
particles ranging from elementary particles and simple atoms to
complex molecular clusters~\cite{CSP09Optics,HGH+12Colloquium}. Most
spectacularly, wave-like behavior has been convincingly demonstrated
even for such large organic compounds as
C$_{60}$[C$_{12}$F$_{25}$]$_{10}$ and
C$_{168}$H$_{94}$F$_{152}$O$_8$N$_4$S$_4$, each comprising 430 atoms
\cite{GET+11Quantum}.

In optics, diffraction is typically portrayed as deflection of light
incident upon an obstacle with sharp boundaries, that can not be
accounted for by reflection or refraction. Interestingly, quantum
mechanics allows for an additional, intrinsically time-dependent
manifestation of the phenomenon: Owing to the dispersive nature of
quantum matter waves, sudden changes in boundary conditions may cause
the particle wave function to develop interference fringes akin to
those in stationary (optical) diffraction problems. This phenomenon,
pioneered in 1952 by Moshinsky \cite{Mos52Diffraction} and presently
referred to as ``diffraction in time'' (DIT), is at the heart of a
vibrant area of experimental and theoretical research concerned with
quantum transients (see Ref.~\cite{CGM09Quantum} for a review).

In the original Moshinsky's setup \cite{Mos52Diffraction}, a
monoenergetic beam of non-relativistic quantum particles is incident
upon a perfectly absorbing barrier (``shutter''). The shutter is
suddenly removed at time $t=t_1$. In other words, the transparency of
the barrier $\chi(t)$, called the aperture function, jumps
instantaneously from 0 at $t<t_1$ to 1 at $t>t_1$, i.e., $\chi(t) =
\Theta(t-t_1)$, with $\Theta$ denoting the Heaviside step
function. The removal of the shutter renders a quantum wave function
with a sharp, discontinuous wave front. The latter disperses in the
course of time, smoothing out the initial discontinuity and developing
a sequence of interference fringes. As Moshinsky has shown, these
fringes bear close mathematical similarity to those in Fresnel
diffraction of light at the edge of a straight, semi-infinite
screen. Moshinsky's paradigmatic triggered considerable interest to
DIT in experimental research with ultra-cold neutrons and atoms, and
Bose-Einstein condensates
\cite{CSP09Optics,HGH+12Colloquium,CGM09Quantum}.

On the theoretical side, a number of interesting extensions and
variations of Moshinsky's shutter problem have been
addressed. Moshinsky himself extended his original result to the case
of a ``time slit'', in which the shutter stays open only during a time
interval $t_1 < t < t_2$, as described by the aperture function
$\chi(t) = \Theta(t-t_1) \Theta(t_2-t)$
\cite{Mos76Diffraction}. Scheitler and Kleber found an exact
analytical solution of a related problem, in which the role of a
smoothly opening shutter was played by a time-dependent
$\delta$-potential barrier $V(x,t) \sim t^{-1} \delta(x)$, with $x$
denoting the spatial coordinate
\cite{SK88adiabaticity,Kle94Exact}. Various physical problems,
requiring generalization of the original Moshinsky's setup to describe
DIT caused by an arbitrary aperture function, $\chi(t)$, have been
addressed in
Refs.~\cite{BZ97Diffraction,CM05Single,CMM07Time,GOC07theorem,TMM+09Enhanced,TMB+11Explanation}. The
analytical methods adopted in all these studies rely on treating the
shutter as an effective ``source'' boundary of a semi-infinite
coordinate space (transmission region). The main advantage of this
approach is that the transmitted wave can be readily expressed in
terms of $\chi(t)$ and a time-dependent source boundary condition. A
significant difficulty with this approach however is that, in general,
there is no unique well-defined recipe for finding the source function
that would accurately mimic a given incident wave packet (although, in
some cases, efficient approximations and exact bounds are known
\cite{GOC07theorem}).

\begin{figure}[ht]
\includegraphics[width=2.5in]{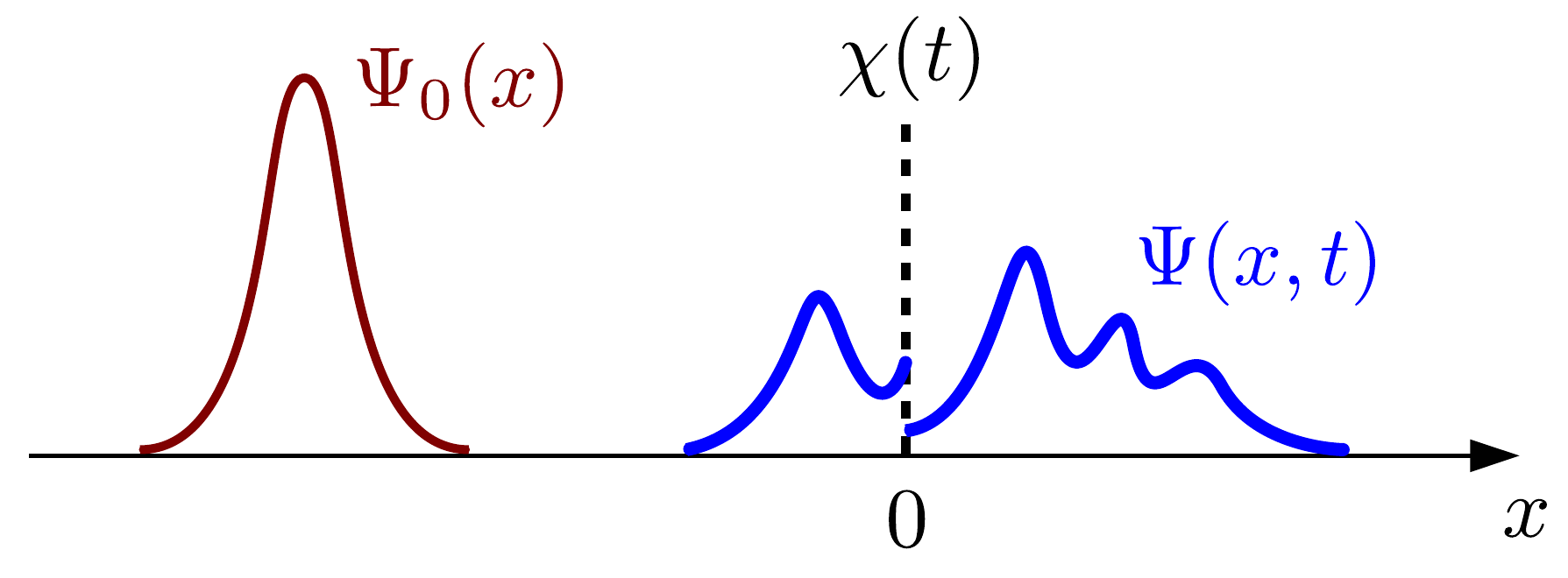}
\caption{Illustration of time-evolution of a quantum state in the
  presence of a time-dependent absorbing shutter.}
\label{fig1}
\end{figure}

In this paper, we introduce a self-consistent {\it exactly solvable}
model of DIT, free of arbitrary parameters. The model aims to describe
dynamics of an {\it arbitrary quantum state} in the presence of an
absorbing time-dependent shutter characterized by an {\it arbitrary
  aperture function} $\chi(t)$, see Fig.~\ref{fig1}. It serves as a
versatile generalization of the original Moshinsky's problem and
reduces to the latter in the particular case $\chi(t) =
\Theta(t-t_1)$.

The central quantity analyzed in this paper is a propagator
$K(x,x';t)$ that relates the particle wave function $\Psi(x,t)$ at
time $t>0$ to an initial quantum state $\Psi_0(x)$ at $t=0$ through
\cite{Bar89Elements}
\begin{equation}
  \Psi(x,t) = \int_{-\infty}^{+\infty} \ud x' \, K(x,x';t) \Psi_0(x')
  \,.
\label{propag-def}
\end{equation}
The propagator satisfies the time-dependent Schr\"{o}dinger equation
(TDSE) on both sides of the absorbing shutter, positioned at $x=0$,
i.e.,
\begin{equation}
  \left( i \partial_t + \frac{\hbar}{2m} \partial_x^2 \right)
  K(x,x';t) = 0 \quad \mathrm{for} \quad x,x' \not= 0 \,.
\label{tdse}
\end{equation}
Here, $m$ denotes the mass of the particle. The propagator is subject
to the initial condition
\begin{equation}
  K(x,x';0^+) = \delta(x-x') \,,
\label{init_cond}
\end{equation}
and is required to vanish as $x \rightarrow \pm \infty$ at negative
imaginary times, i.e., at $t = -i |t|$.

Our treatment of absorption is based on a time domain rendition of an
approach originally introduced by Kottler. The approach applies to
diffraction of stationary fields governed by the Poisson's equation at
spatial apertures in otherwise perfectly absorbing (``black'') screens
\cite{Kot23Zur,Kot65Diffraction}. According to this approach, a wave
originating on one side of a black screen is subject to a
discontinuous jump at every point of the screen, except for points
inside possible openings (holes, slits, etc.). More concretely, the
difference of the wave amplitude on the source side of the screen and
that on the opposite side equals the amplitude of the corresponding
free-space wave, i.e., the amplitude the wave would have if no screen
were present. A similar discontinuity condition is imposed on the
normal derivative of the field. Both the field and its normal
derivative are postulated continuous across the openings. As Kottler
has shown, an exact solution of the Poisson's equation, subject to the
above well-defined, though unusual, discontinuous boundary conditions,
is identical to the wave field predicted by Kirchhoff's diffraction
theory.

In our model, we consider a time-dependent generalization of the
Kottler's discontinuity condition at the shutter. Thus, we require the
propagator to satisfy
\begin{equation}
  K(x,x';t) \big|_{x=0^-}^{x=0^+} = \sgn(x') \big[ 1-\chi(t) \big]
  K_0(x-x',t) \big|_{x=0}
\label{discont1}
\end{equation}
and
\begin{align}
 \partial_x K(x,x'; &t) \big|_{x=0^-}^{x=0^+} \nonumber\\ &= \sgn(x')
 \big[ 1-\chi(t) \big] \partial_x K_0(x-x',t) \big|_{x=0}
 \,. \label{discont2}
\end{align}
Here, $\chi(t)$ denotes the aperture function, ranging between 0
(shutter closed) and 1 (shutter open),
\begin{equation}
  K_0(z,t) = \sqrt{\frac{m}{2 \pi i \hbar t}} \exp \left( i \frac{m}{2
    \hbar t} z^2 \right)
\label{free_prop}
\end{equation}
is the free-particle propagator, and $\sgn(x') = x'/|x'|$ stands for
the sign function. The latter identifies whether the source is located
to the left or right of the shutter.

Equations~(\ref{tdse}--\ref{discont2}) completely specify dynamics of
a quantum particle in the presence of a time-dependent absorbing
barrier. In fact, as will be shown below, this quantum-mechanical
problem admits an exact analytic solution valid for an arbitrary
piecewise differentiable function $\chi(t)$. The solution is given by
\begin{equation}
  K(x,x';t) = \Xi(x,x') K_0(x-x',t) + K_1(x,x';t)
\label{K_sol1a}
\end{equation}
with
\begin{equation}
  K_1(x,x';t) = \int_0^t \ud \tau \, u \, K_0(x,t-\tau) \chi(\tau)
  K_0(-x',\tau)
\label{K_sol1b}
\end{equation}
and
\begin{equation}
  u(x,x';t,\tau) = -\frac{\sgn(x')}{2} \left( \frac{x}{t-\tau} -
  \frac{x'}{\tau} \right) \,.
\label{K_sol1c}
\end{equation}
Here, $\Xi(x,x') = [1 + \sgn(x) \sgn(x')]/2$, i.e., $\Xi$ equals 1 if
$x$ and $x'$ lie on the same side of the barrier, and 0 otherwise.

For what follows below, it is useful to rewrite the propagator, given
by Eqs.~(\ref{K_sol1a}--\ref{K_sol1c}), in an alternative
form. Evaluating the integral in Eq.~(\ref{K_sol1b}) by parts, we
obtain
\begin{equation}
  K(x,x';t) = \Xi(x,x') \big[ 1-\chi(t) \big] K_0(x-x',t) +
  K_2(x,x';t) \,,
\label{K_sol2a}
\end{equation}
where
\begin{align}
  K_2&(x,x';t) = \frac{1}{2} \bigg( \chi(0) + \chi(t) \nonumber\\ &+
  \sgn(x') \int_0^t \ud \tau \, \frac{\ud \chi(\tau)}{\ud \tau} \,
  \erf(\Phi) \bigg) K_0(x-x',t)
\label{K_sol2b}
\end{align}
and
\begin{equation}
  \Phi(x,x';t,\tau) = \sqrt{\frac{m}{2 i \hbar t}} \left( x
  \sqrt{\frac{\tau}{t-\tau}} + x' \sqrt{\frac{t-\tau}{\tau}} \right)
  \,.
\label{Phi}
\end{equation}

Equations~(\ref{K_sol2a}--\ref{Phi}), while being mathematically
equivalent to Eqs.~(\ref{K_sol1a}--\ref{K_sol1c}), enable a
straightforward verification of the fact that $K$ satisfies
Eqs.~(\ref{tdse}--\ref{discont2}). Indeed, since both $K_2$ and
$\partial_x K_2$ are continuous functions of $x$, the validity of the
discontinuity conditions, Eqs.~(\ref{discont1}) and (\ref{discont2}),
follows directly from Eq.~(\ref{K_sol2a}). The validity of the initial
condition, Eq.~(\ref{init_cond}), can be also verified
straightforwardly: $K(x,x';0^+) = [\Xi + (1-\Xi) \chi(0)] K_0
(x,x';0^+) = \delta(x-x')$. The fact that $K$ satisfies the TDSE is
confirmed by a direct substitution of Eqs.~(\ref{K_sol2a}--\ref{Phi})
into Eq.~(\ref{tdse}). (Here, it is convenient to take into account
the identity $\partial_t \Phi + \frac{x-x'}{t} \partial_x \Phi + i
\frac{\hbar}{m} \Phi (\partial_x \Phi)^2 = 0$.) Finally, we note that
uniqueness of the solution can be established in a standard way
\cite{Bar89Elements}.

It is important to point out that, due to nonunitarity of quantum
evolution in the presence of absorption, the propagator $K$, in
general, does not fulfill the composition property, i.e.,
\begin{equation}
  K(x,x';t) \not= \int_{-\infty}^{+\infty} \ud \xi \, K(x,\xi;t-\tau)
  K(\xi,x';\tau) \,.
\label{no_comp_prop}
\end{equation}
This can be readily seen by considering the simple case of a
time-independent, completely absorbing barrier, for which $\chi = 0$
and $K = \Xi K_0$.

Although not valid generally, the composition property holds in the
following important special case. Consider a scenario, in which the
absorbing barrier acts only up to some time $t_{\mathrm{f}}$. In other
words, suppose the aperture function $\chi$ is such that $\chi(\tau) =
1$ for all $\tau > t_{\mathrm{f}}$. It can then be shown that $K$
satisfies
\begin{equation}
  K(x,x';t) = \int_{-\infty}^{\infty} \ud \xi \, K_0(x-\xi,t-\tau)
  K(\xi,x',\tau)
\label{comp_prop-spec_case}
\end{equation}
for $0 < t_{\mathrm{f}} < \tau < t$. The physical picture offered by
Eq.~(\ref{comp_prop-spec_case}) is apparent: Once the absorbing
shutter has been switched off, the wave function evolves in accordance
with the free-particle propagator.

Equations~(\ref{K_sol1a}--\ref{K_sol1c}) offer the following physical
interpretation of the wave function evolution. The full propagator $K$
is a sum of $\Xi K_0$, describing propagation in the case of a
completely absorbing barrier ($\chi = 0)$, and $K_1$, representing
contribution of a barrier of nonzero transparency. In the transmission
region, i.e., for $x$ and $x'$ such that $\Xi(x,x')=0$, the expression
for $K_1$, Eq.~(\ref{K_sol1b}), conforms to the Huygens-Fresnel
principle \cite{Gou12Huygens}: The probability amplitude at the point
$x$ and time $t$, produced by a point source at $x'$ and time 0, can
be viewed as that produced by a fictitious source, located at the
origin (between $x$ and $x'$); the strength of the fictitious source
is determined by the free-particle wave function at the origin,
modulated by the aperture function.

In certain cases, the integral in Eq.~(\ref{K_sol2b}) can be evaluated
explicitly. One important example is that of a ``time grating'',
characterized by a ``staircase'' aperture function $\chi(\tau) =
\chi_0 \Theta(t_1-\tau) + \sum_{n=1}^{N-1} \chi_n \Theta(\tau-t_n)
\Theta(t_{n+1}-\tau) + \chi_N \Theta(\tau-t_N)$, where $0 \leq \chi_j
\leq 1$, with $0 \leq j \leq N$, and $0 < t_1 < t_2 < \ldots < t_N <
t$. In this case, the integral in Eq.~(\ref{K_sol2b}) reduces to
$\sum_{n=1}^N (\chi_{n+1} - \chi_n) \erf [ \Phi(x,x';t;t_n) ]$,
rendering a closed form expression for the propagator. In particular,
$\chi(t) = \Theta(t-t_1)$ leads to a propagator coincident with that
in the original Moshinsky's shutter problem \cite{Gou12Huygens}.

\begin{figure*}[ht]
\includegraphics[width=6.5in]{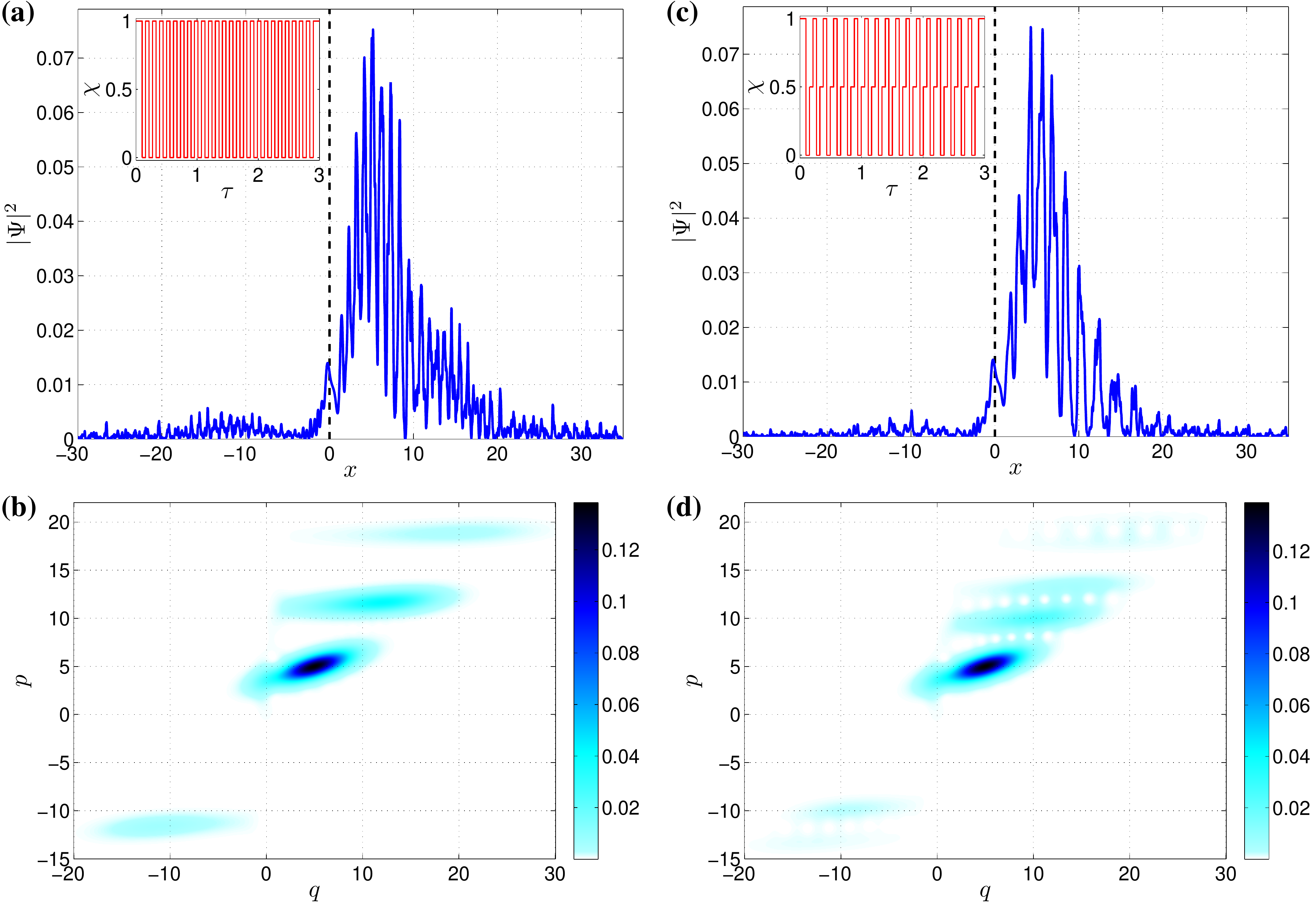}
\caption{Diffraction in time of an initially Gaussian wave
  packet. Figures (a) and (b) show, respectively, the probability
  density and Husimi representation of the diffracted wave packet for
  the aperture function given in the inset of (a). Similarly, figures
  (c) and (d) correspond to the aperture function given in the inset
  of (c). All quantities are given in atomic units, $m = \hbar = 1$.}
\label{fig2}
\end{figure*}

A diffracted wave function, $\Psi(x,t)$, resulting from an arbitrary
initial state, $\Psi_0(x)$, can now be obtained by numerically
evaluating the integral in Eq.~(\ref{propag-def}). As an illustration,
we consider diffraction of an initially localized Gaussian wave
packet. Adopting atomic units, $m = \hbar = 1$, we consider a coherent
state $\psi_{q,p}(x) = (\pi)^{-1/4} \exp [ -\frac{1}{2} (x-q)^2 + i p
  (x-q) ]$, with the average position $q$ and momentum $p$, and set
the initial state to be $\Psi_0(x) = \psi_{-10, \,
  5}(x)$. Figure~\ref{fig2} shows the wave function $\Psi(x,t)$ at
time $t=3$ diffracted at two different time gratings $\chi(\tau)$,
shown in the insets of Figs.~\ref{fig2}a and \ref{fig2}c. The
elementary, one-period cell of the first grating, Fig.~\ref{fig2}a, is
given by $0 \times \Theta(\tau) \Theta(\Delta t - \tau) + 1 \times
\Theta(\tau - \Delta t) \Theta(2 \Delta t - \tau)$, and that of the
second grating, Fig.~\ref{fig2}c, by $0 \times \Theta(\tau)
\Theta(\Delta t - \tau) + 0.5 \times \Theta(\tau - \Delta t) \Theta(2
\Delta t - \tau) + 1 \times \Theta(\tau - 2 \Delta t) \Theta(3 \Delta
t - \tau)$, with $\Delta t = 0.056$. (The value $\chi=0.5$ means that
the shutter allows only a half of the probability amplitude through,
while absorbing the other half.) The probability densities
$|\Psi(x,t)|^2$ for the two gratings are shown in the main panels of
Figs.~\ref{fig2}a and \ref{fig2}c, and the corresponding Husimi
distributions, $H(q,p) = \left| \int_{-\infty}^{+\infty} \ud x \,
\psi_{q,p}^* (x) \Psi(x,t) \right|^2$, with the asterisk denoting
complex conjugation, are presented in Figs.~\ref{fig2}b and
\ref{fig2}d, respectively. (Phase-space representations, akin to
$H(q,p)$, have been previously used for establishing classical-quantum
analogies in the dynamics of diffracted particles
\cite{MMS99Diffraction}.) The central bright peaks in
Figs.~\ref{fig2}b and \ref{fig2}d are localized around the point
$(q,p)=(5,5)$, which is the average phase-space location of the
particle in the absence of a shutter. Additionally, in the case of the
first grating, three separated diffraction peaks -- one in the
reflection region, $q < 0$, and two in the transmission region, $q >
0$, -- are clearly visible in Fig.~\ref{fig2}b. On the contrary, in
the case of the second grating, Fig.~\ref{fig2}d, diffraction peaks
overlap substantially, forming a distinct pattern of probability
naughts, i.e., points in phase space ``avoided'' by the diffracted
particle. We point out that, in general, the diffraction pattern
sensitively depends on the initial wave packet and aperture function,
making diffraction on time gratings well suited for spectroscopic
analysis of quantum states.

So far, we have shown that Eqs.~(\ref{K_sol1a}--\ref{K_sol1c}), or
equivalently Eqs.~(\ref{K_sol2a}--\ref{Phi}), give an exact solution
to a quantum dynamical problem, defined by
Eqs.~(\ref{tdse}--\ref{discont2}). A natural question is however in
order: Does this dynamical system accurately model physical reality?
In what follows, we argue that the proposed system should indeed
provide a good description of the motion of a quantum particle in the
presence of a purely absorbing (but not reflecting, in the classical
sense) barrier. The key building block of our model is a
time-dependent extension of the Kottler's discontinuity conditions,
Eqs.~(\ref{discont1}) and (\ref{discont2}). The latter, in the case of
stationary waves governed by the Poisson's equation, are
mathematically equivalent to Kirchhoff's description of
diffraction. Predictions of the Kirchhoff's theory, in turn, are
generally found in good agreement with experimental data in the
transmission region, while somewhat lacking accuracy in the reflection
region \cite{NHL95Diffraction}. Therefore, at the very least, it is
reasonable to expect the propagator $K(x,x';t)$ to correctly describe
quantum wave dynamics in the transmission region. Moreover, if the
absorbing shutter is in effect only until some final time
$t_{\mathrm{f}}$, so that $\chi(t) = 1$ for all $t > t_{\mathrm{f}}$,
then $K(x,x';t)$, being an analytic function of $x$ at any time
$t>t_{\mathrm{f}}$, must correctly describe the dynamics of the
diffracted particle in both the transmission and the reflection
regions.

Nevertheless, in the absence of a compelling local theory of
absorption, the ultimate answer to the question of how accurately the
proposed model describes physical reality can only be given by an
experimental investigation. To this end, atom-optics systems appear to
be particularly suitable. Remarkably, recent progress in control and
manipulation of ultra-cold atoms has made it possible to perform
diffraction and interference experiments with a single, isolated atom,
corresponding to a quantum wave packet highly localized in space
\cite{PHB12Observation}. At the same time, strong ionizing radiation
has been used to realize an optical diffraction grating, similar in
effect to an absorbing nanostructure (mechanical) grating
\cite{HDG+13universal}. Utilizing these technologies, one might
envision a single-atom DIT experiment, in which quasi-one-dimensional
motion of an atom is ``intercepted'' by a time-dependent absorbing
shutter produced by a transversely-oriented, pulsing sheet of ionizing
radiation.

In conclusion, we have addressed a self-consistent mathematical model
of diffraction in time and found its exact analytical solution in form
of a time-dependent propagator. The latter enables a quantitative
description of diffraction of an arbitrary initial quantum state at an
absorbing shutter governed by an arbitrary aperture function. We
believe that our model will prove useful in the areas of coherent
quantum control, quantum metrology, in designing new diffraction and
interference experiments with atoms and molecules, and, more
generally, in exploring foundations of quantum physics.

\acknowledgements

The author thanks Ilya Arakelyan, Orestis Georgiou, and Jonathan
M. Robbins for their valuable comments.

%%%%%%%%%%%%%%%%%%%%%%%%%%%%%%%%%%%%%%%%%%%%%%%%%%%%%%%%%%%%%%%%%%%%%%%%%%%%%%%%
%\bibliographystyle{naturemag}
%\bibliography{../diffraction-interference,../mystuff}

\end{document}